\documentclass[aps,prb,preprint,groupedaddress,showpacs,amsmath,amssymb]{revtex4}

\usepackage{graphicx}
\usepackage{dcolumn}
\usepackage{bm}

\begin{document}

\title{Thermal expansion of CaFe$_2$As$_2$: effect of cobalt doping and post-growth thermal treatment}

\author{Sergey L. Bud'ko, Sheng Ran, and Paul C. Canfield}
\affiliation{Ames Laboratory, US DOE and Department of Physics and Astronomy, Iowa State University, Ames, IA 50011, USA}

\date{\today}

\begin{abstract}

We report thermal expansion measurements on Ca(Fe$_{1-x}$Co$_{x}$)$_2$As$_2$ single crystals with different thermal treatment, with samples chosen to represent four different ground states observed in this family. For all samples thermal expansion is anisotropic with different signs of the in-plane and $c$-axis thermal expansion coefficients in the high temperature, tetragonal phase.  The features in thermal expansion associated with the phase transitions are of opposite  signs as well, pointing to a different response of transition temperatures to the in-plane and the $c$-axis stress. These features, and consequently the inferred pressure derivatives, are very large, clearly and substantially exceeding those in the  Ba(Fe$_{1-x}$Co$_{x}$)$_2$As$_2$ family.  For all transitions the $c$-axis response is dominant. 
\end{abstract}

\pacs{74.70.Xa, 65.40.De, 74.25.Dw, 75.30.Kz}

\maketitle

\section{Introduction}
Of many iron-arsenide based superconductors, \cite{ish09a,joh10a,ste11a,joh11a} members of the AFe$_2$As$_2$ (A = Ba, Sr, Ca, Eu, ...) \cite{can10a,nin11a,man10a} family have become model systems to study magnetism, superconductivity and their interplay in these new superconductors. In the AFe$_2$As$_2$ family, and most probably throughout all iron-arsenide based superconductors and their parent compounds, CaFe$_2$As$_2$  (i) provides an extreme example of the strongly coupled structural and magnetic phase transitions \cite{nin08a,can09a}, and (ii) demonstrates  extreme sensitivity to the pressure or stress. \cite{can09a,tor08a,kre08a,gol09a,yuw09a,lee09a,tor09a,pro10a} This extreme sensitivity to pressure allows a sample to be tuned from an antiferromagnetic, orthorhombic ground state to non-moment bearing, collapsed tetragonal one by applying just few kilobars of pressure. \cite{can09a} Additionally, if a non-hydrostatic pressure component is present in the experiment, a (in many cases filamentary) superconducting phase can be detected \cite{tor08a,lee09a,tor09a,pro10a,par08a}, that is probably associated with a non-collapsed  tetragonal phase being stabilized as part of a mixture of several crystallographic phases in a CaFe$_2$As$_2$ sample at low temperatures. \cite{pro10a}.

Recently, the collapsed tetragonal phase and bulk superconductivity were induced in CaFe$_2$As$_2$ at ambient pressure by manipulation of nanoscale precipitates through judicious annealing and quenching, combined with cobalt substitution in place of iron. \cite{ran11a,ran12a} The effects of annealing and quenching were shown to be similar to those of the pressure. \cite{ran11a,ran12a,gat12a}  It was suggested that the annealing and quenching process, via the precipitates, controls the amount of strain built up in the samples and this way acts similarly to a moderate applied external pressure. It was shown \cite{ran12a} that by careful choice of Co - concentration and annealing and quenching parameters, the  Ca(Fe$_{1-x}$Co$_x$)$_2$As$_2$ system can by systematically and reproducibly tuned to have one of four different ground states: orthorhombic and antiferromagnetic, tetragonal and superconducting, tetragonal and non-superconducting, and collapsed tetragonal without magnetic order or superconductivity. The equivalence between annealing and quenching and pressure was explored further in Ref. [\onlinecite{gat12a}], where three different low temperature phases were observed in one sample on application of a moderate hydrostatic pressure, below 3 kbar, and record-high in their absolute values, for inorganic compounds, pressure derivatives of the N\'eel temperature ($T_N$), superconducting transition temperature ($T_c$), and structural transition temperature to the collapsed tetragonal phase ($T_{cT}$), were measured.

Thermal expansion measurements were performed on pure, Sn - grown, CaFe$_2$As$_2$ \cite{bud10a} and several other iron-arsenide based superconductors. Exceptionally large, negative thermal expansion was recently observed in La - doped  CaFe$_2$As$_2$. \cite{reb12a} Carefully controlled annealing and quenching in the  Ca(Fe$_{1-x}$Co$_x$)$_2$As$_2$ system opens opportunities for the measurements that are either impossible or difficult to perform under pressure. In addition, such samples offer a clear comparison to the Ba(Fe$_{1-x}$Co$_x$)$_2$As$_2$ materials (with their greatly reduced, or more usual, pressure sensitivities). \cite{col09a} In this work we report thermal expansion measurements on a number of  Ca(Fe$_{1-x}$Co$_x$)$_2$As$_2$ samples that cover all salient parts of the three-dimensional $T - x - T_{anneal/quench}$ phase diagram from Ref. [\onlinecite{ran12a}] with different low temperature states.  Generally speaking, thermal expansion measurements in this system reveal the evolution of the phase transitions with doping and thermal treatment and allow for the evaluation of (inferred through the Ehrenfest and Clausius - Clapeyron relations for the second and first order phase transitions, respectively) anisotropic, uniaxial pressure derivatives of different phase transitions observed in this system. The latter is of particular interest given the reported, large overall pressure dependences and high sensitivity to non-hydrostaticity for these samples.

\section{Experimental details}
The single crystals of Ca(Fe$_{1-x}$Co$_x$)$_2$As$_2$  were grown out of a FeAs/CoAs flux using conventional high temperature growth techniques. Details of growth, annealing and quenching procedures as well as basic physical properties are described in extensive detail in Refs. [\onlinecite{ran11a,ran12a}]. The actual cobalt concentrations were determined using a wavelength-dispersive x-ray spectroscopy, these experimental concentration values are used in the text.  The samples used in this work are either from exactly the batches studied in Refs. [\onlinecite{ran11a,ran12a}], or from very similarly grown and thermally treated batches.

Thermal expansion data were obtained using a capacitive dilatometer constructed of OFHC copper; a detailed description of the dilatometer is presented elsewhere. \cite{sch06a} The dilatometer was mounted in a Quantum Design Physical Property Measurement System, PPMS-14, instrument and was operated over a temperature range of 1.8 - 300 K. Due to the morphology of the crystals, for most of the samples the dilation was measured only along the $c$-axis. For selected samples the dilation in the $ab$-plane was measured as well. In those cases arbitrary in-plane orientations were used. The data were collected on warming. The heat capacity was measured using a hybrid adiabatic relaxation technique of the heat capacity option in a Quantum Design, PPMS-14 instrument.

The choice of visually uniform samples and using only mild polishing and small forces in the dilatometer allowed us to perform measurements through the strongly first order structural transitions without destruction of the samples. In some cases though the first order transition was observed as two, closely spaced, sharp steps in thermal dilation (shown as two symbols for the same concentration in the $x - T$ phase diagram, see below, inset to Fig. \ref{F1}(b)).

\section{Results} 

\subsection{As grown Ca(Fe$_{1-x}$Co$_x$)$_2$As$_2$}

The data in Ref. [\onlinecite{ran12a}] suggest that as-grown  ($T_{anneal/quench} = 960^\circ$ C) Ca(Fe$_{1-x}$Co$_x$)$_2$As$_2$ ($x \leq 0.059$) samples all have transition from room-temperature tetragonal to low-temperature collapsed tetragonal phase.  Magnetization measurements show the $T_{cT}$ value to be practically temperature-independent, with the width of the transition increasing with Co-concentration.  Resistance measurements suffered from cracking of the samples on going through the structural transition for $x \leq 0.22$ and had no discernible features for $0.28 \leq x \leq 0.059$. \cite{ran12a}. 

\begin{figure}
\begin{center}
\includegraphics[angle=0,width=100mm]{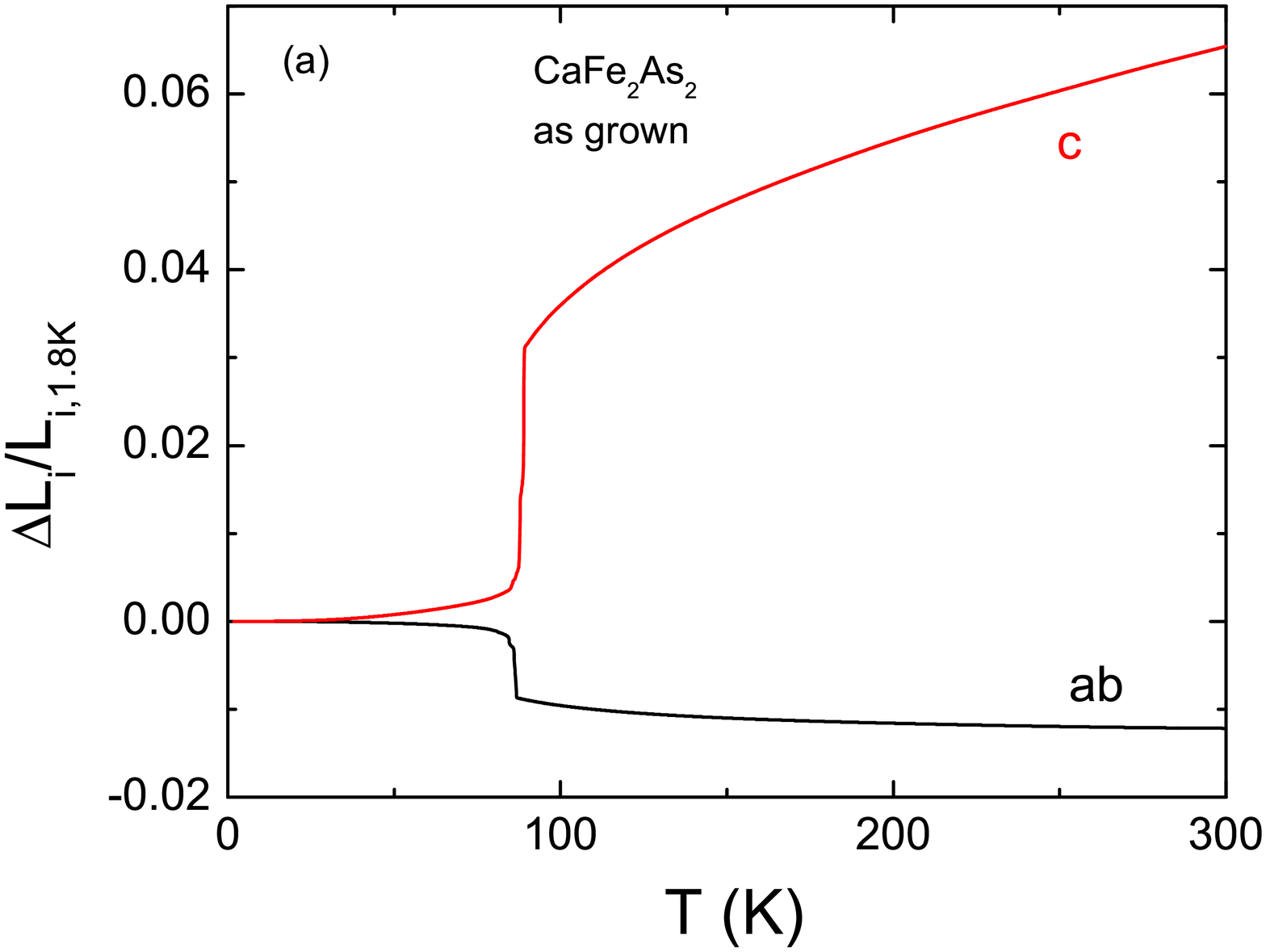}
\includegraphics[angle=0,width=100mm]{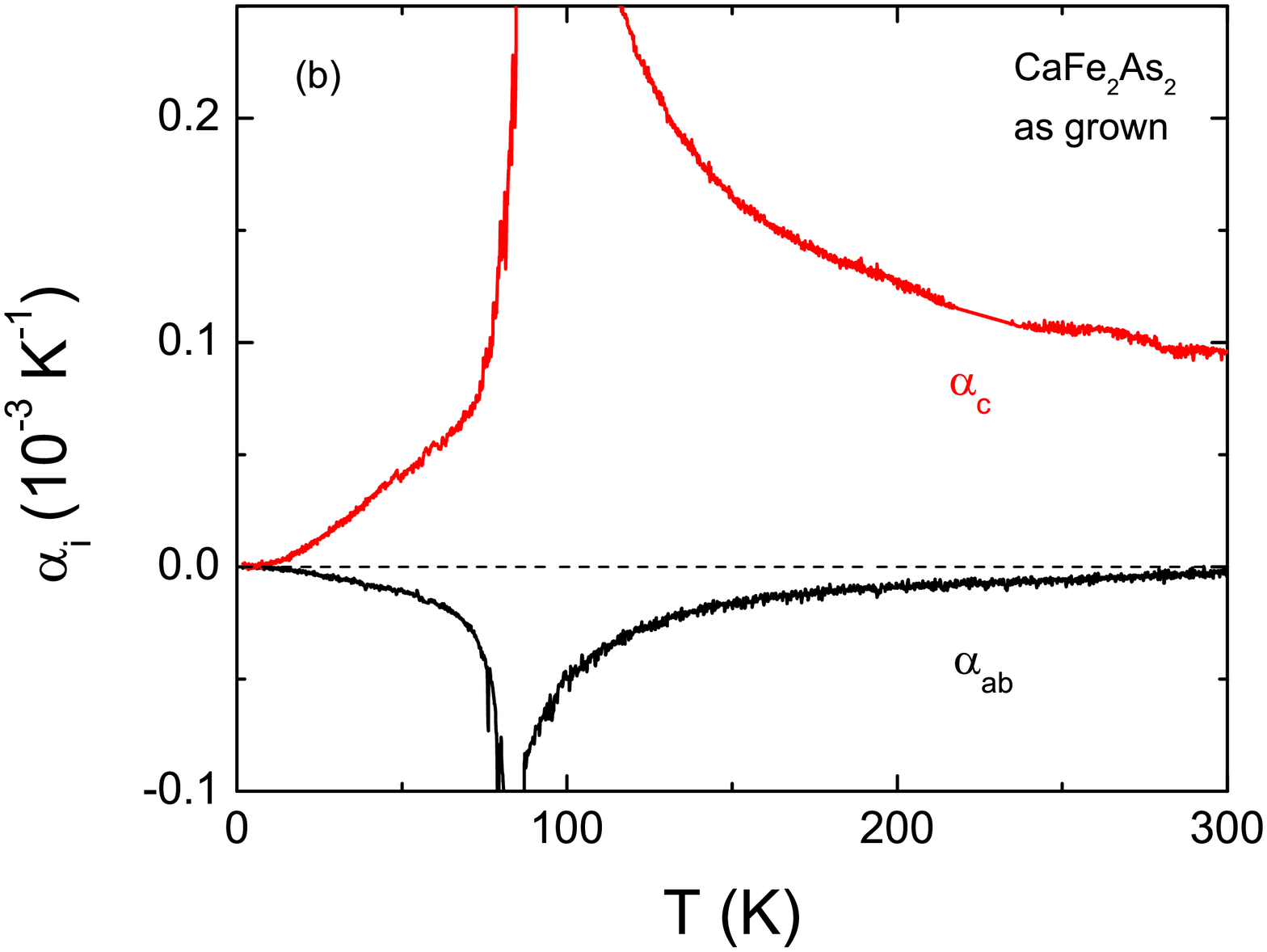}
\end{center}
\caption{(Color online) (a) Temperature-dependent, anisotropic dilation with respect to 1.8 K ($\Delta L_i/L_{i, 1.8~\textrm{K}} = [L_i(T) - L_i(1.8$ K$)]/L_i(1.8$ K$)$) of as-grown CaFe$_2$As$_2$; (b) temperature-dependent, anisotropic thermal expansion coefficients, $\alpha_i$,  of as-grown CaFe$_2$As$_2$.} \label{F2}
\end{figure}

Temperature-dependent $c$-axis dilation data of as-grown Ca(Fe$_{1-x}$Co$_x$)$_2$As$_2$ ($x \leq 0.059$) are shown in Fig. \ref{F1}(a). The overall relative change of the $c$-axis lengths appears to be essentially Co-concentration independent, amounting to  $[c(1.8~$K$) - c(300~$K$)]/c(300~$K$) \sim - 0.06$, of which 30 - 40 \% being associated with the transition to the collapsed tetragonal phase. The sharpness of the transitions, at least for lower Co - concentrations, is consistent with them being the first order. As consistent with with the magnetization measurements, \cite{ran12a} the transition temperature does not change significantly with Co - substitution, but the width of the transition increases (Fig. \ref{F2}(b)). The $c$-axis thermal expansion coefficients above the structural transition to the collapsed tetragonal phase are fairly independent of the cobalt concentration.  Results for $x = 0$ and $x = 0.059$ obtained using temperature-dependent single crystal x-ray diffraction \cite{ran12a} are fairly consistent with the dilation data presented here.

For as grown  ($T_{anneal/quench} = 960^\circ$ C), pure CaFe$_2$As$_2$, anisotropic thermal expansion was measured (Fig. \ref{F2}). The thermal expansion is very anisotropic, over the whole temperature range (1.8 - 300 K) in-plane thermal expansion coefficient, $\alpha_{ab}$, is negative and $c$-axis coefficient, $\alpha_c$, is positive. There is a significant change of the lattice parameters on the transition from the tetragonal to the collapsed tetragonal phase: the $a$ - lattice parameter increases by $\sim 0.75 \%$ and $c$ lattice parameter decreases by $\sim 2.8 \%$. These values as well as the temperature dependences of the lattice parameters are consistent with the single crystal x-ray diffraction data. \cite{ran12a}

\subsection{Ca(Fe$_{0.977}$Co$_{0.023}$)$_2$As$_2$ with different thermal treatment}

\begin{figure}
\begin{center}
\includegraphics[angle=0,width=120mm]{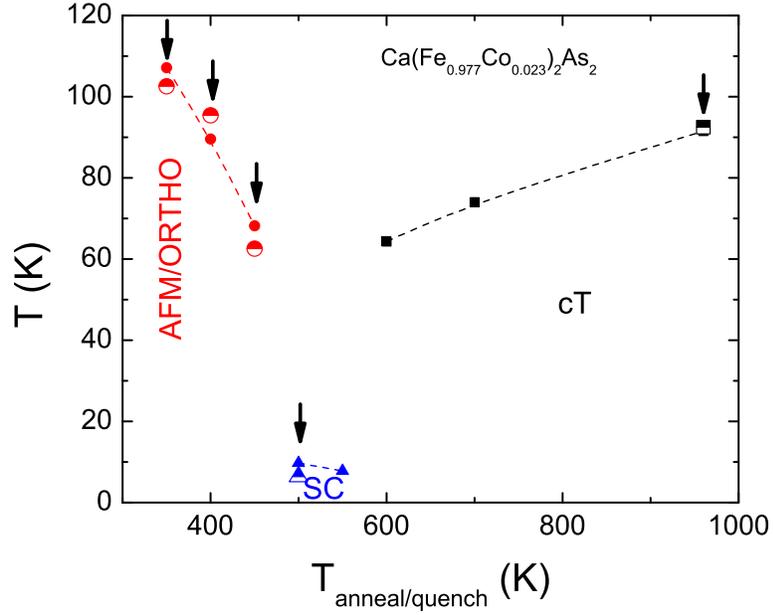}
\end{center}
\caption{(Color online) $T - T_{anneal/quench}$ phase diagram for Ca(Fe$_{0.977}$Co$_{0.023}$)$_2$As$_2$ (after Ref. [\onlinecite{ran12a}], filled symbols). AFM/ORTHO, SC, and cT label antiferromagnetic and orthorhombic, tetragonal and superconducting, and non-moment bearing collapsed tetragonal phases, respectively. Arrows mark the concentrations used in this work and transition temperatures determined from the thermal expansion data are shown as half - filled symbols.} \label{F3}
\end{figure}

The second set of samples chosen for the $c$-axis thermal expansion measurements is the Ca(Fe$_{0.977}$Co$_{0.023}$)$_2$As$_2$ that is annealed for 7 days at different temperatures, from 350$^\circ$ C to 960$^\circ$ C and then quenched (note, that 960$ ^\circ$ C annealing refers to as grown sample, see  Refs. [\onlinecite{ran11a,ran12a}]). This cut of the three-dimensional $T - x - T_{anneal/quench}$ phase diagram allows for access to three different ground states: antiferromagnetic and orthorhombic, tetragonal and superconducting, and non-moment bearing collapsed tetragonal (Fig. \ref{F3}).

\begin{figure}
\begin{center}
\includegraphics[angle=0,width=100mm]{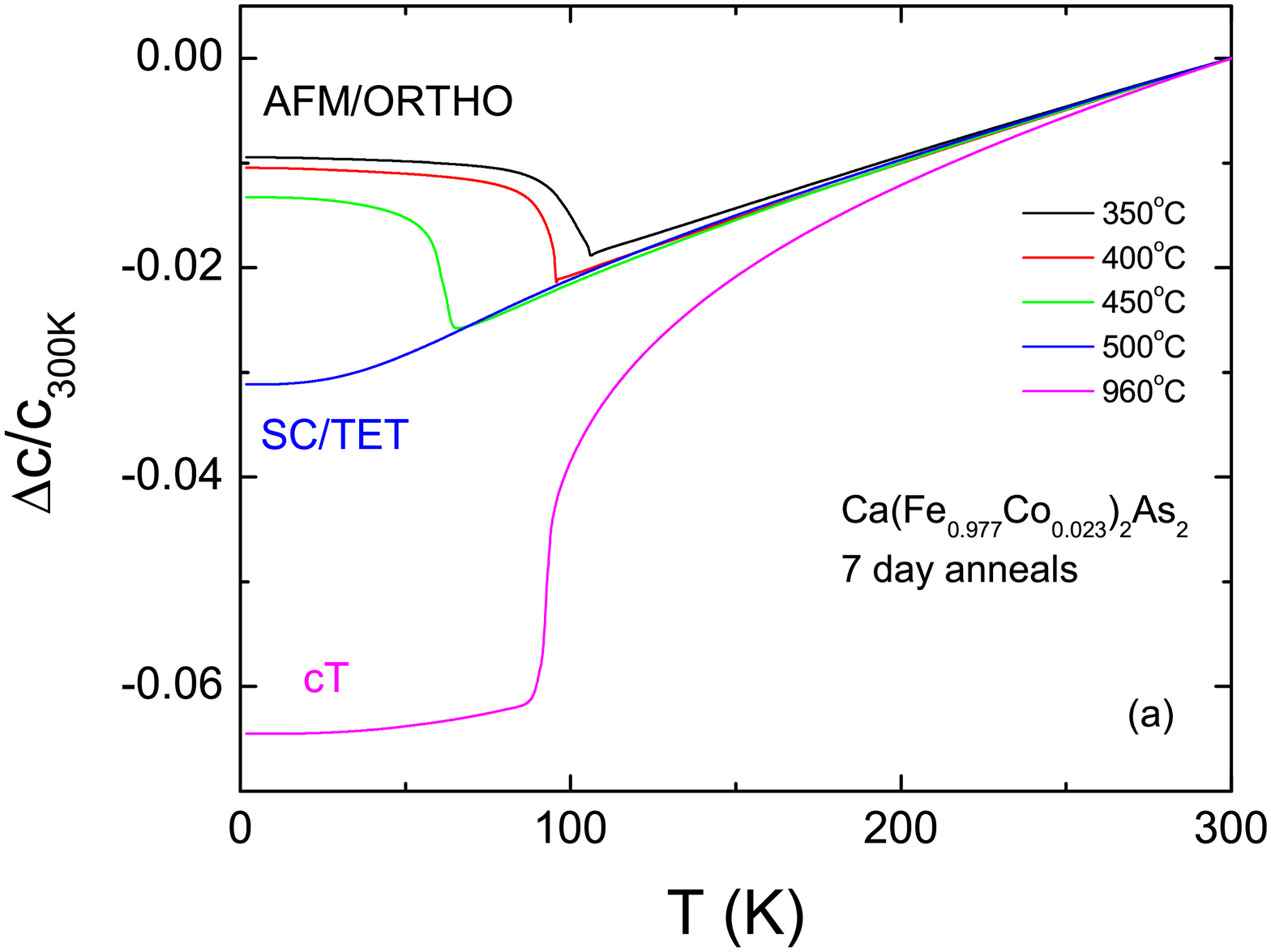}
\includegraphics[angle=0,width=100mm]{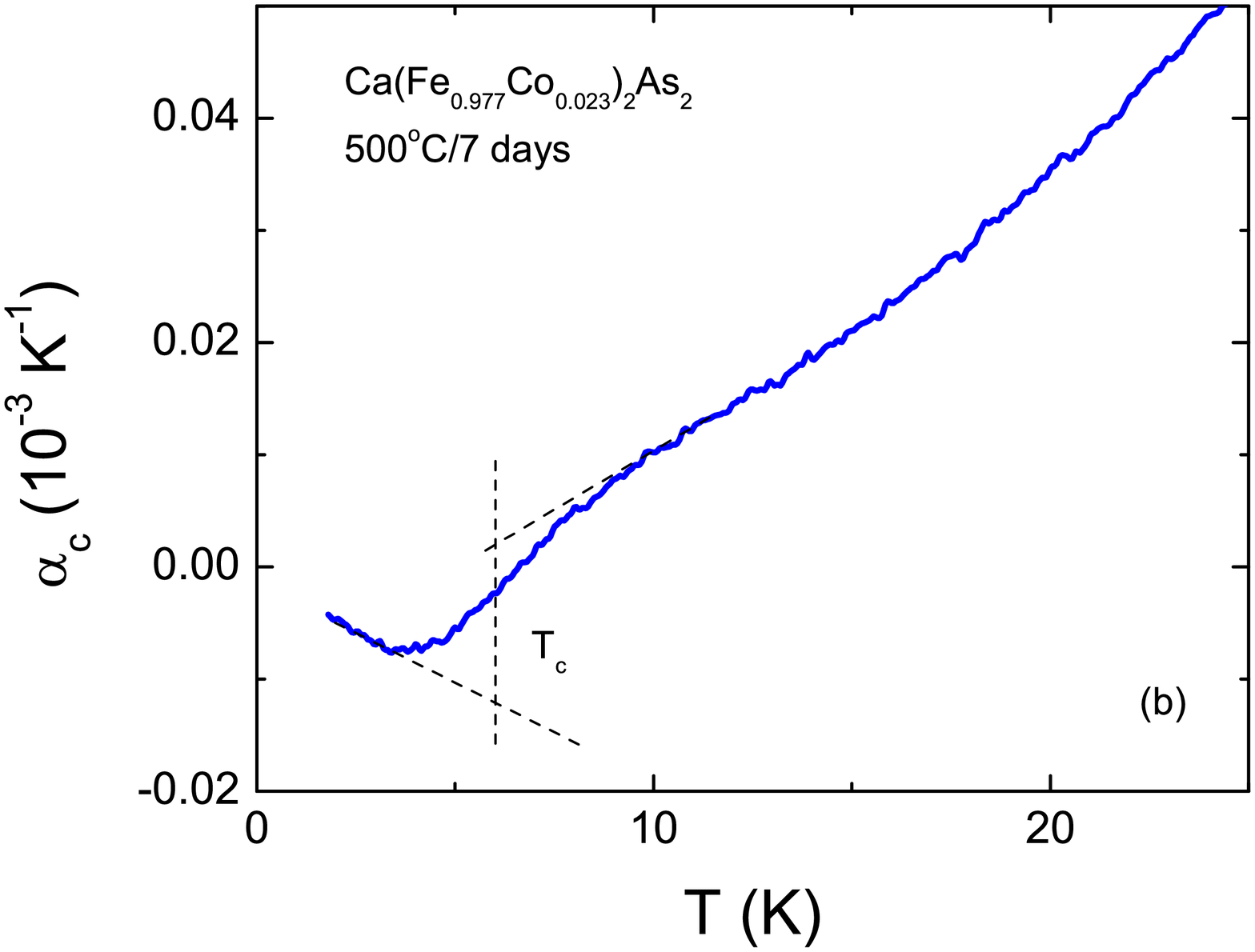}
\end{center}
\caption{(Color online) (a) Temperature-dependent, anisotropic dilation  with respect to 300 K ($\Delta c/c_{300~\textrm{K}} = [c(T) - c(300$ K$)]/c(300$ K$)$) of Ca(Fe$_{0.977}$Co$_{0.023}$)$_2$As$_2$ with different thermal treatment. AFM/ORTHO, SC/TET, and cT label antiferromagnetic and orthorhombic, tetragonal and superconducting, and collapsed tetragonal and paramagnetic phases, respectively. (b) Low temperature, $c$-axis thermal expansion coefficient, $\alpha_c$,  of Ca(Fe$_{0.977}$Co$_{0.023}$)$_2$As$_2$ annealed for 7 days at 500$^\circ$ C. $T_c$ is approximately determined using the same criterion as in Ref. [\onlinecite{bud09a}].} \label{F4}
\end{figure}

Fig. \ref{F4}(a) presents $c$-axis, temperature-dependent, dilation of  Ca(Fe$_{0.977}$Co$_{0.023}$)$_2$As$_2$ samples with different thermal treatments. The transitions from tetragonal to antiferromagnetic and orthorhombic phase, and from tetragonal to collapsed tetragonal phase are clearly seen in the $c$-axis dilation. The $c$-axis length for these two transitions changes in the opposite way: it sharply increases at the former transition and decreases, even more dramatically, at the latter. The antiferromagnetic and structural transition temperature decreases with increase of the annealing temperature, in accordance to the data in Ref. [\onlinecite{ran12a}], there is no evidence of a split between antiferromagnetic and structural transitions, and the change in the $c$-axis lattice parameter at this transition is rather large, $\sim 1 - 1.5 \%$. The overall change of the $c$-axis between 300 K and 1.8 K is around $ - 1\%$ for samples with antiferromagnetic and orthorhombic ground state, $\sim -3\%$ for the superconducting tetragonal sample and $\sim -6\%$ for the sample with the collapsed tetragonal ground state.  The superconducting transition is clearly seen (as a low temperature minimum) in the $c$-axis thermal expansion coefficient (Fig. \ref{F4}b).

\subsection{Ca(Fe$_{1-x}$Co$_x$)$_2$As$_2$, $x \leq 0.058$, annealed at 400$^\circ$ C for 7 days}

\begin{figure}
\begin{center}
\includegraphics[angle=0,width=120mm]{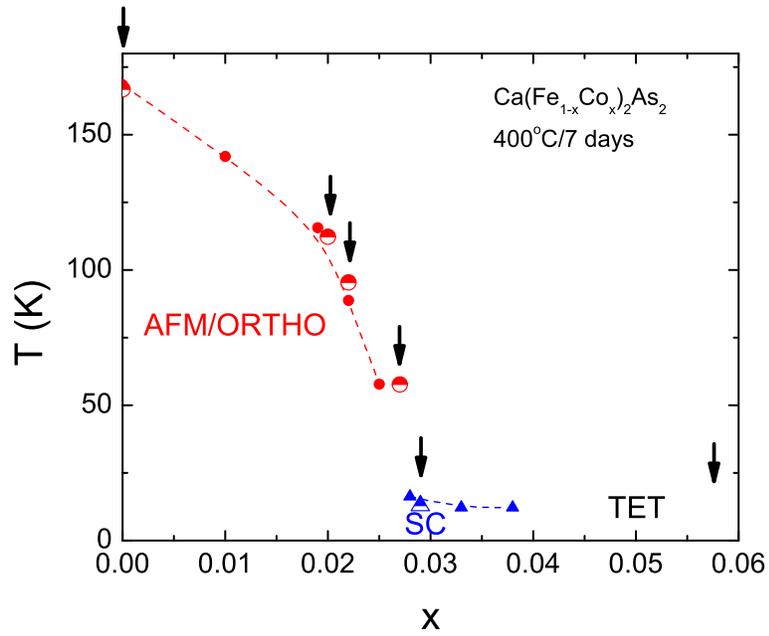}
\end{center}
\caption{(Color online) $T - x$ phase diagram for Ca(Fe$_{1-x}$Co$_x$)$_2$As$_2$,  $x \leq 0.058$, annealed at 400$^\circ$ C for 7 days  (after Ref. [\onlinecite{ran12a}], filled symbols). AFM/ORTHO, SC, and TET label antiferromagnetic and orthorhombic, tetragonal and superconducting, and tetragonal and paramagnetic phases, respectively. Arrows mark the concentrations used in this work and transition temperatures determined from the thermal expansion data are shown as half - filled symbols.} \label{F5}
\end{figure}

\begin{figure}
\begin{center}
\includegraphics[angle=0,width=100mm]{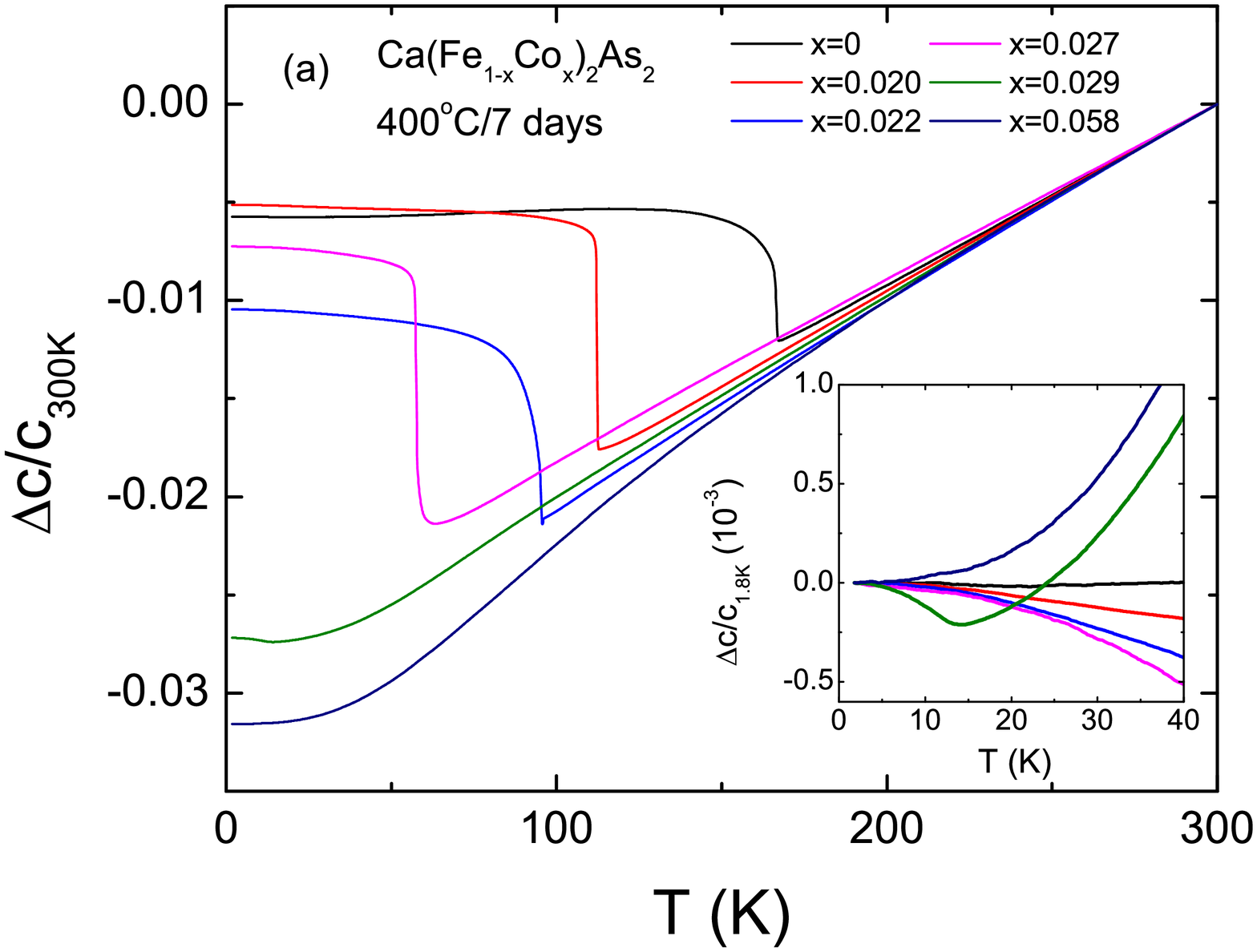}
\includegraphics[angle=0,width=100mm]{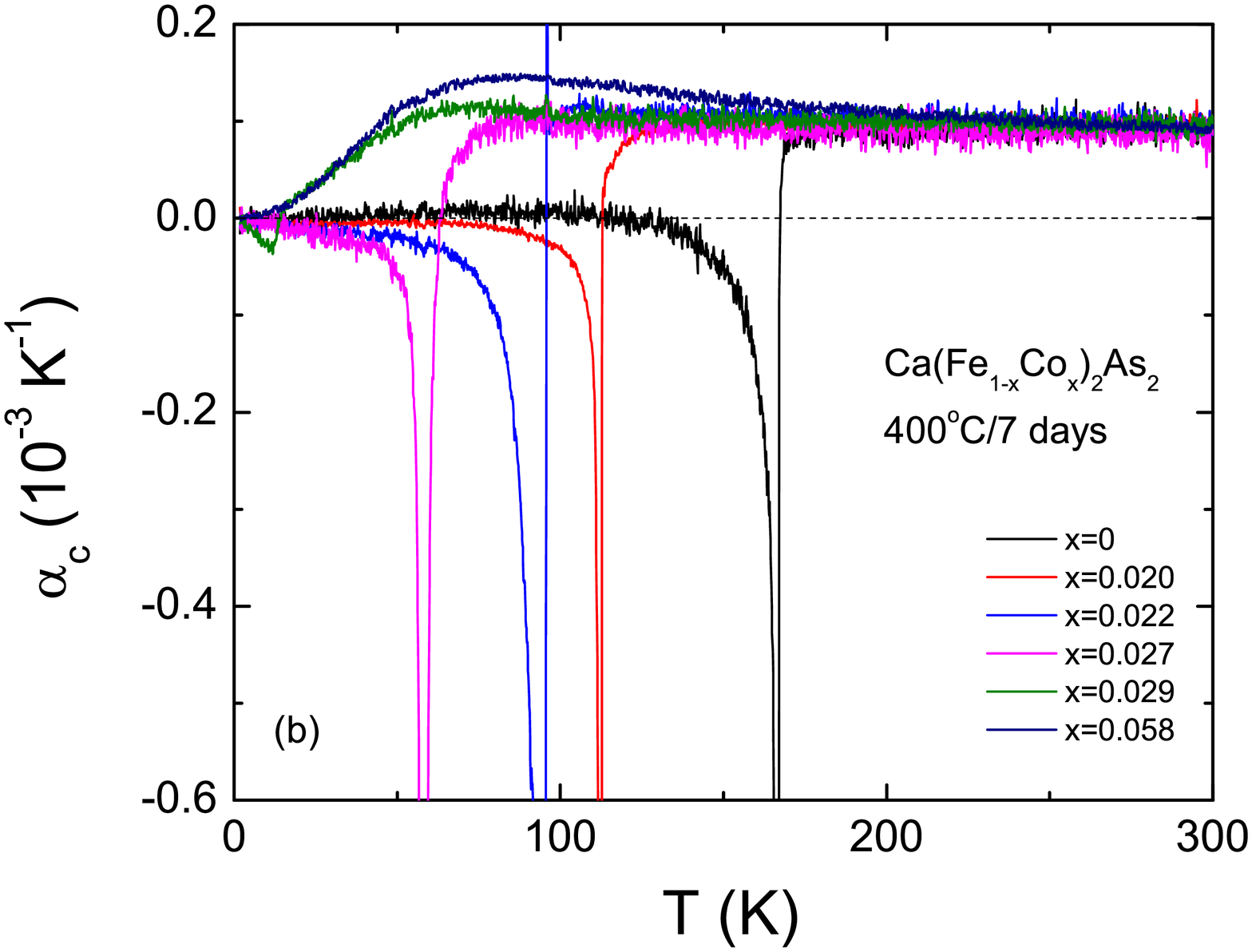}
\end{center}
\caption{(Color online) (a) Temperature-dependent, $c$-axis dilation with respect to 300 K ($\Delta c/c{300~\textrm{K}} = [c(T) - c(300$ K$)]/c(300$ K$)$) of Ca(Fe$_{1-x}$Co$_x$)$_2$As$_2$, $x \leq 0.058$, annealed at 400$^\circ$ C for 7 days. Inset: low temperature part of the data plotted with respect to the values at 1.8 K.  (b) Temperature-dependent, $c$-axis thermal expansion coefficients, $\alpha_c$,  for the same samples.} \label{F6}
\end{figure}

With the Ca(Fe$_{1-x}$Co$_x$)$_2$As$_2$ ($x \leq 0.058$) samples annealed at 400$^\circ$ C for 7 days and then quenched we can access antiferromagnetic and orthorhombic, tetragonal and superconducting, and tetragonal and paramagnetic (with no superconductivity or long range magnetic order) ground states (Fig. \ref{F5}). Temperature-dependent $c$-axis dilation and $c$-axis thermal expansion coefficient for these samples are shown in Fig. \ref{F6}. Step-like features associated with coupled antiferromagnetic and structural transition are clearly seen for the samples with $x \leq 0.027$ in both dilation and thermal expansion coefficient. The sharpness of the transition is consistent with it being the first order. Given that the data for $x = 0$ are comparable with the results for Sn flux grown CaFe$_2$As$_2$, \cite{bud10a} we conclude that small, coherent, sub-micron crystallites of Fe-As \cite{ran11a} appear to have no significant effect on the $c$-axis thermal expansion. The characteristic temperatures of these features are consistent with the published phase diagram. \cite{ran12a}  A change in the, otherwise smooth, temperature-dependent $c$-axis dilation and thermal expansion coefficient of the Ca(Fe$_{0.971}$Co$_{0.029}$)$_2$As$_2$ sample at about 15 K is a signature of the superconducting transition. This clear feature strongly suggests that superconductivity is bulk. Apart from this feature, samples with $x = 0.029$ and $x = 0.058$ behave essentially the same way.

\begin{figure}
\begin{center}
\includegraphics[angle=0,width=100mm]{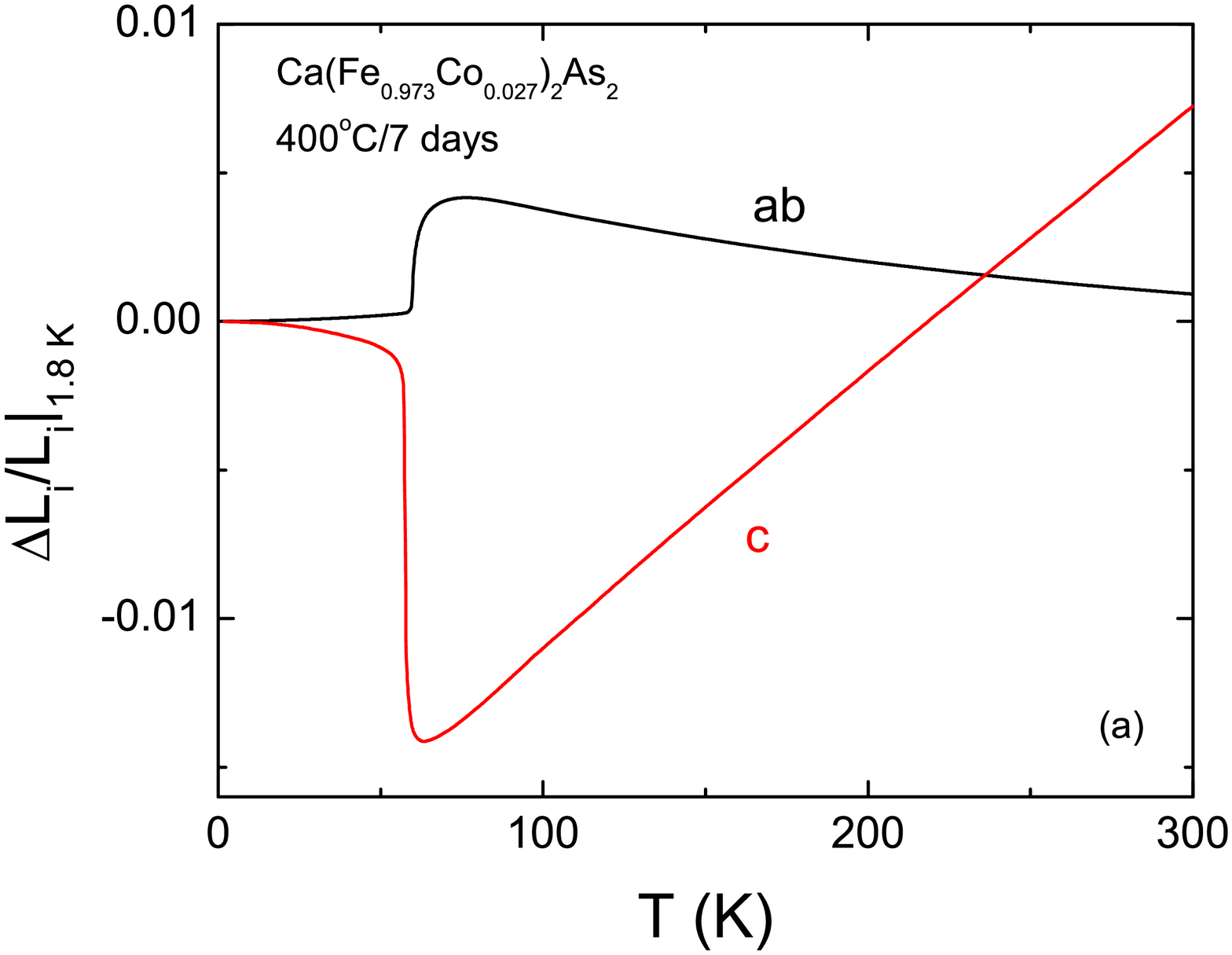}
\includegraphics[angle=0,width=100mm]{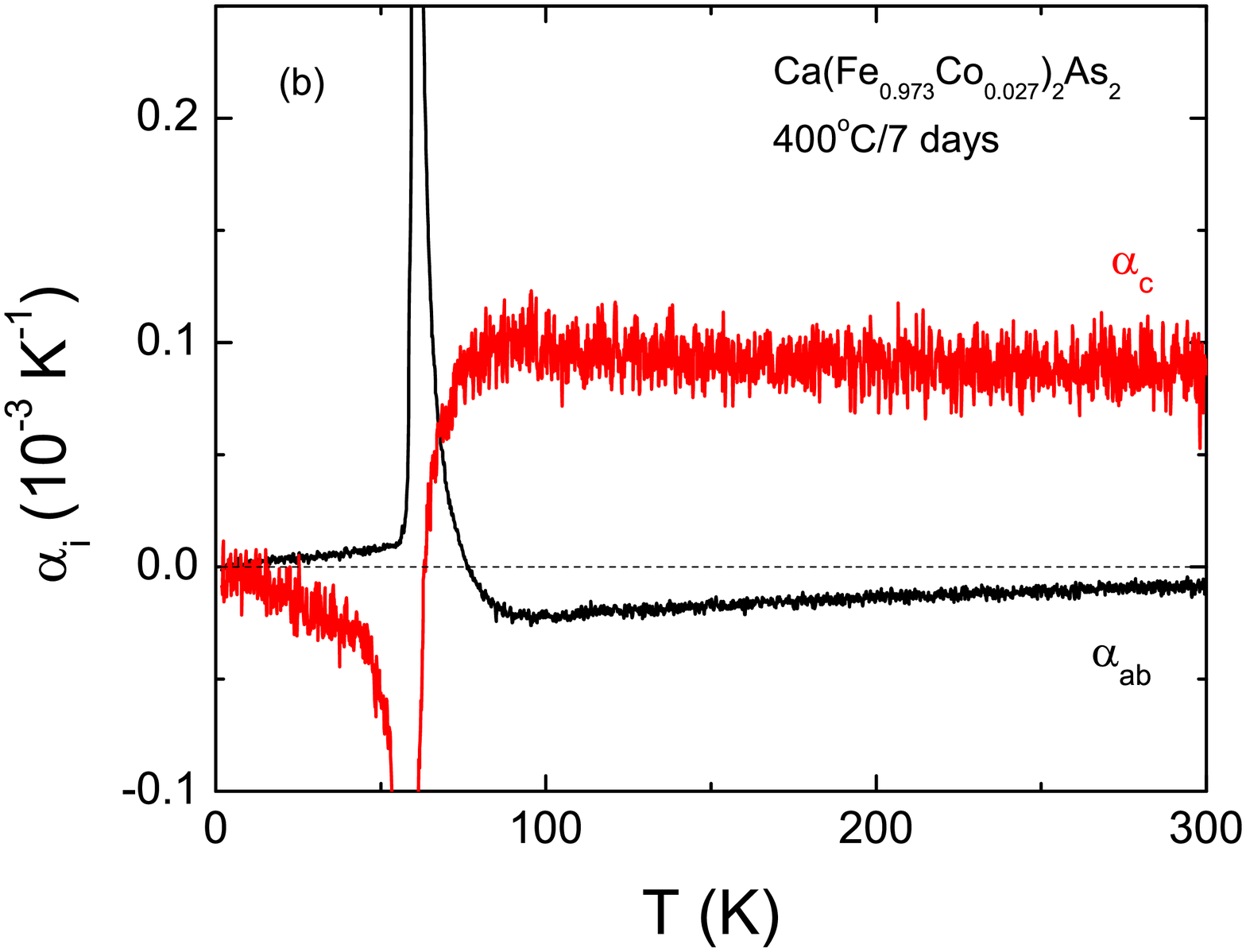}
\end{center}
\caption{(Color online) (a) Temperature-dependent, anisotropic dilation with respect to 1.8 K ($\Delta L_i/L_{i, 1.8~\textrm{K}} = [L_i(T) - L_i(1.8$ K$)]/L_i(1.8$ K$)$) of Ca(Fe$_{0.973}$Co$_{0.027}$)$_2$As$_2$, annealed at 400$^\circ$ C for 7 days; (b) temperature-dependent, anisotropic thermal expansion coefficients, $\alpha_i$,  of the same sample.} \label{F7}
\end{figure}

\begin{figure}
\begin{center}
\includegraphics[angle=0,width=100mm]{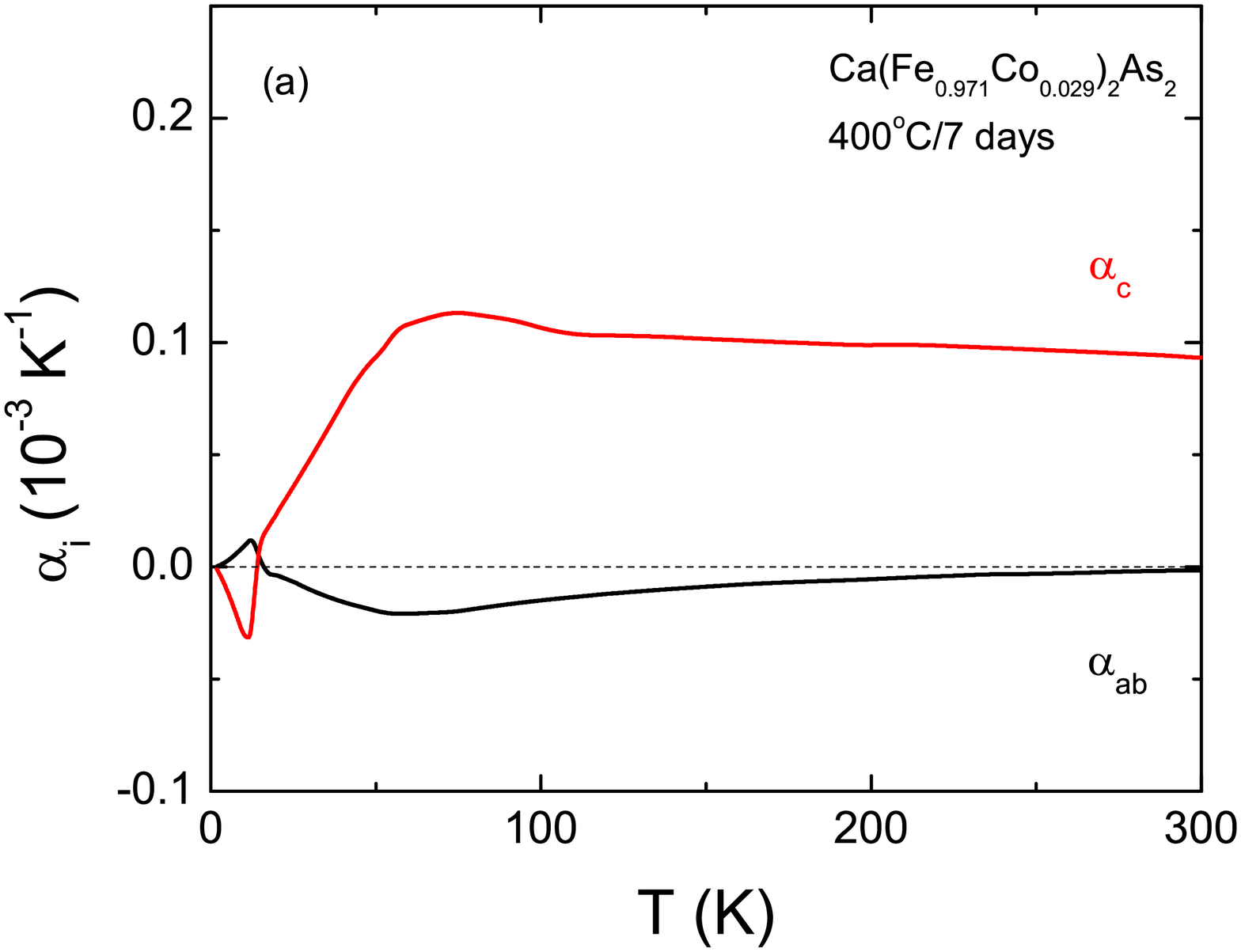}
\includegraphics[angle=0,width=100mm]{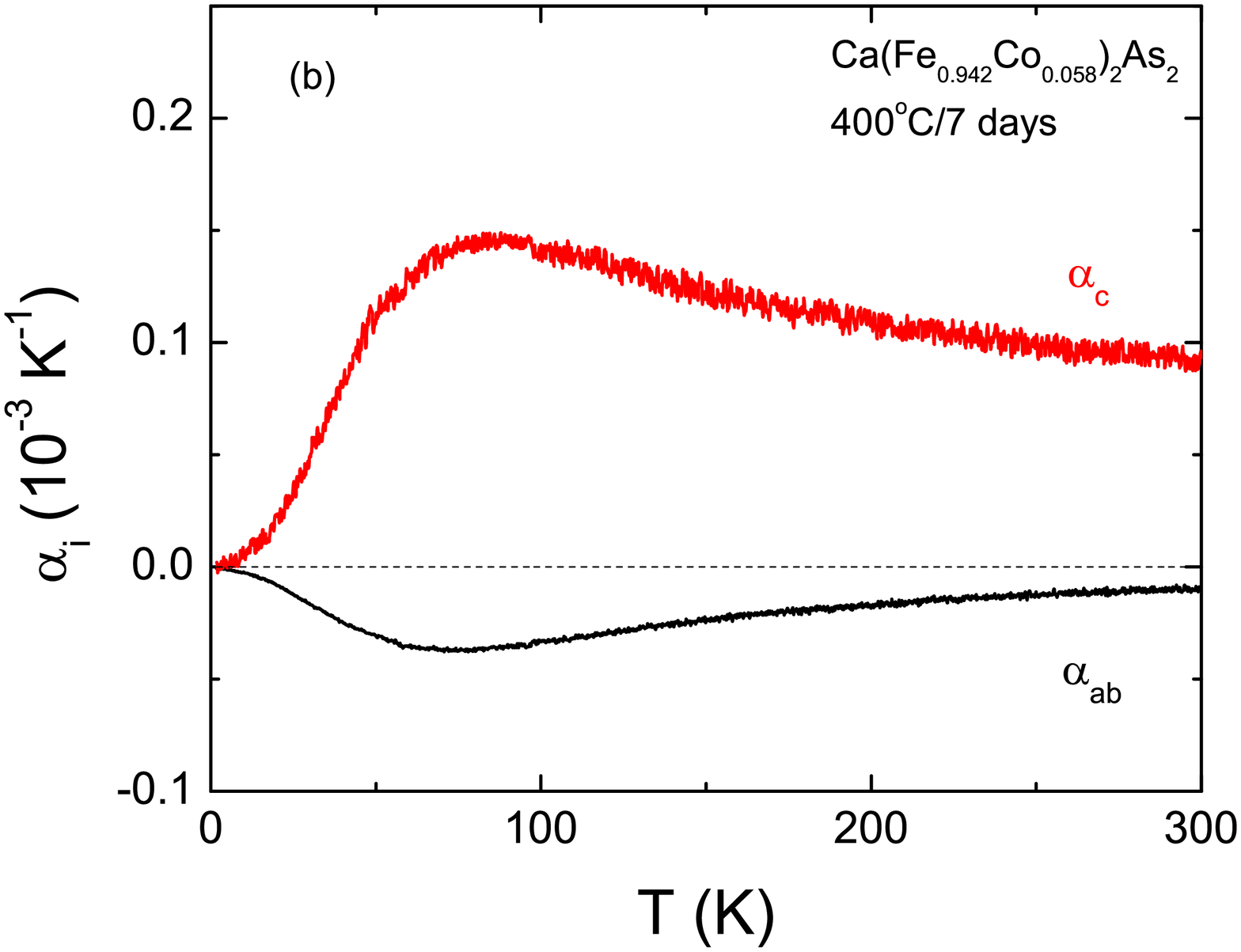}
\end{center}
\caption{(Color online) Temperature-dependent, anisotropic thermal expansion coefficients, $\alpha_i$,  of (a)  Ca(Fe$_{0.971}$Co$_{0.029}$)$_2$As$_2$, annealed at 400$^\circ$ C for 7 days; (b)  Ca(Fe$_{0.942}$Co$_{0.058}$)$_2$As$_2$, annealed at 400$^\circ$ C for 7 days.} \label{F8}
\end{figure}

For selected samples, with Co concentration of  $x = 0.027$, $x = 0.029$, and $x = 0.058$, annealed at 400$^\circ$ C for 7 days, anisotropic thermal expansion measurements were performed. The results are shown in Figs. \ref{F7}, \ref{F8}. For all three samples thermal expansion is very anisotropic, with in-plane thermal expansion coefficient being negative, and $c$-axis coefficient being positive in the tetragonal phase (e.g. above antiferromagnetic and structural, or superconducting transition). 

For  $x = 0.027$ sample the sharp jumps in thermal dilation associated with the antiferromagnetic and structural transition are of opposite signs. At the transition from tetragonal to antiferromagnetic and orthorhombic phase, the average in-plane lattice parameter decreases by $\sim 0.4 \%$ and the $c$-axis increases by $\sim 1.4 \%$. (Note that due to existence of domains in the antiferromagnetic and orthorhombic phase, for the in-plane direction we measure an approximate average of $a$ and $b$ lattice parameters.) These changes are significant and comparable  (albeit smaller) in absolute values to the changes at transition to the collapsed tetragonal phase, but are of the opposite signs (compare Fig. \ref{F2}(a) and Fig. \ref{F7}(a)).

Thermal expansion coefficients of the samples with $x = 0.029$ and $x = 0.058$ (Fig. \ref{F8}) are very similar, except for the clear signatures of superconducting transition in the former. These features are of the opposite sign for in-plane and $c$-axis directions. The apparent difference in noise levels in the data for these two samples is probably caused by difference in the samples' dimensions and by digital differentiation procedure.

\section{Discussion and summary}

For all samples studied in this work, thermal expansion in the tetragonal phase (above the phase transitions) is very similar, with negative thermal expansion coefficients in the $ab$ - plane and positive, and of larger absolute value, along the $c$-axis, resulting in positive volume thermal expansion coefficient. All phase transitions have large, clear features in thermal expansion. No discernible splitting of the structural and antiferromagnetic transitions was observed.

\begin{figure}
\begin{center}
\includegraphics[angle=0,width=100mm]{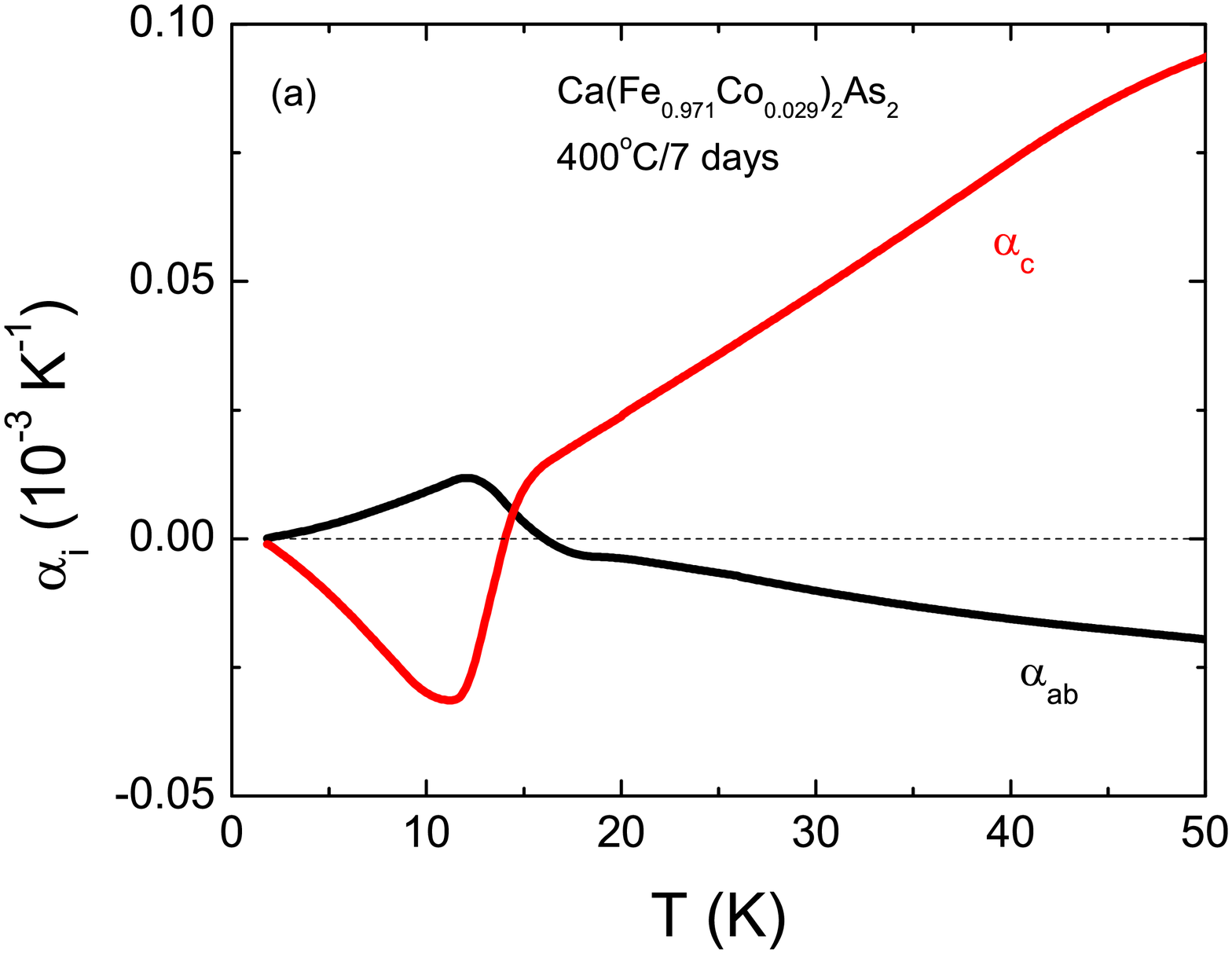}
\includegraphics[angle=0,width=100mm]{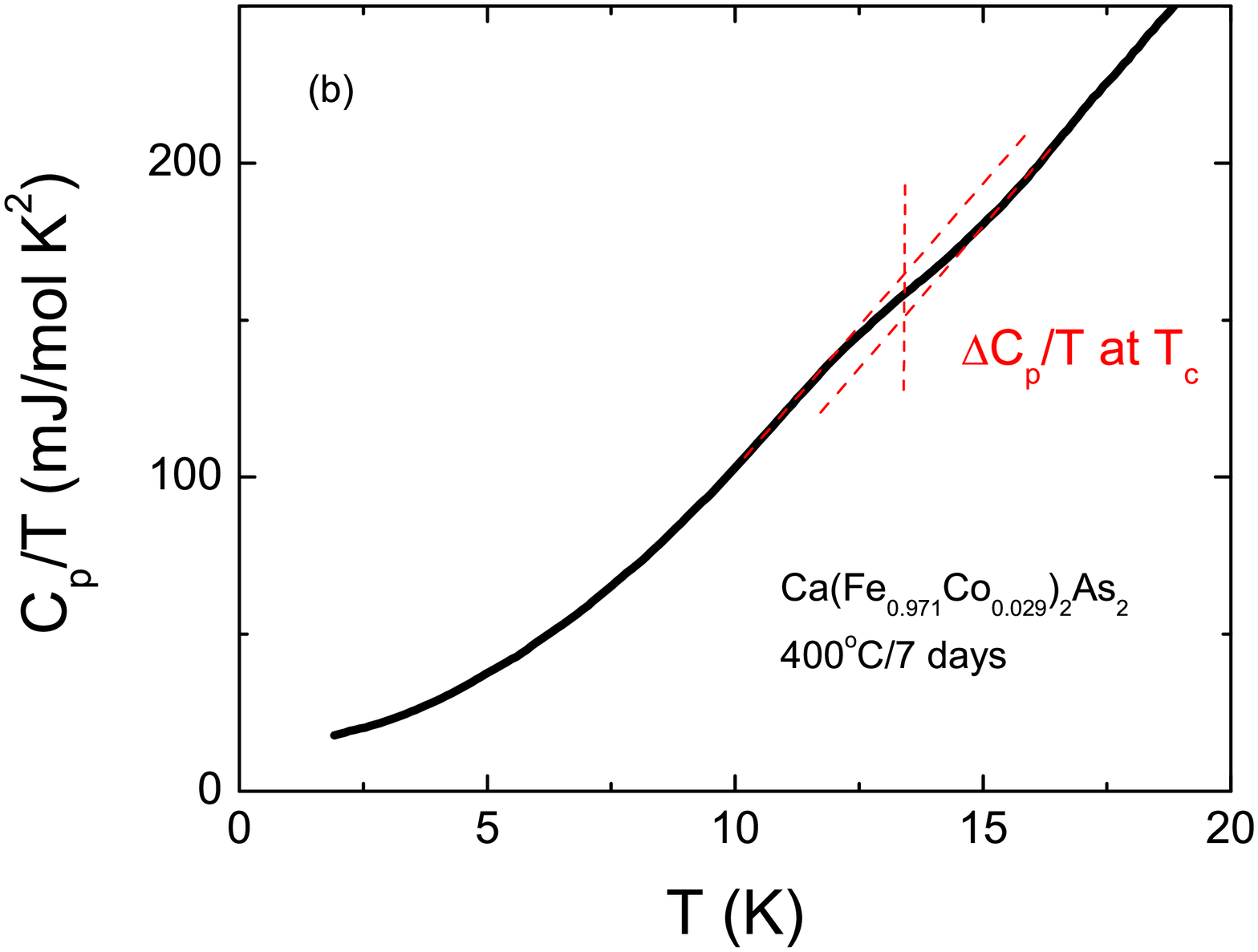}
\end{center}
\caption{(Color online) (a) Temperature-dependent, anisotropic thermal expansion coefficients, $\alpha_i$,  of  Ca(Fe$_{0.971}$Co$_{0.029}$)$_2$As$_2$, annealed at 400$^\circ$ C for 7 days below 50 K; (b)  Heat capacity, plotted as $C_p/T$ of  Ca(Fe$_{0.971}$Co$_{0.029}$)$_2$As$_2$, annealed at 400$^\circ$ C for 7 days below 50 K.} \label{F9}
\end{figure}

The superconducting transition observed in Ca(Fe$_{0.971}$Co$_{0.029}$)$_2$As$_2$,  annealed at 400$^\circ$ C for 7 days, is a second order phase transition, and we can infer the uniaxial pressure derivatives of $T_c$ using the Ehrenfest relation, \cite{bar99a}   $dT_c/dp_i = \frac{V_m \Delta \alpha_i T_c}{\Delta C_p}$, where $V_m$ is the molar volume ($V_m \approx 5.4 \cdot 10^{-5}$ m$^3$ for CaFe$_2$As$_2$, using the lattice parameters \cite{nin08a} at 300 K, can serve as a reasonable approximation), $\Delta \alpha_i~ (i = a, c)$ is the jump in the thermal expansion coefficient at the phase transition, and $\Delta C_p$ is the corresponding jump in the heat capacity. Anisotropic thermal expansion coefficients and heat capacity at low temperatures, near the superconducting transition for $x = 0.029$ sample are shown in Fig. \ref{F9}. From these data the uniaxial pressure derivatives can be inferred: $d T_c/d p_a \approx 7$ K/kbar, and $d T_c/d p_c \approx - 20$ K/kbar, resulting a rough estimate of the hydrostatic pressure derivative, $d T_c/dP \approx 2 \cdot d T_c/dp_a + d T_c/d p_c \approx - 6$ K/kbar. This $d T_c/dP$ value is close to that directly measured under hydrostatic pressure. \cite{gat12a} The  $d T_c/d p_c \approx - 20$ K/kbar value is phenomenally large and provides further evidence of the extreme pressure and stress sensitivity of these compounds.

Transitions from tetragonal to collapsed tetragonal phase and from tetragonal to antiferromagnetic and orthorhombic phase are believed to be the first order. In this case Clausius - Clapeyron relation \cite{bar99a} can be used to infer the uniaxiail pressure dependences of the critical temperatures: $dT_{crit}/d p_i = \frac{V_m \Delta L_i/L_i}{\Delta S}$, where $\Delta L_i/L_i$ is a relative change of length in the $i$-th direction at the transition and $\Delta S$ is the corresponding change of entropy. Of these parameters, $\Delta S$ is difficult to measure with a reasonable precision. To circumvent this, we will use the experimentally measured values of hydrostatic $d T_{crit}^{exp}/d P$ (from the literature) and the relative change of volume at the transition (from this work, taken as  $\Delta V/V \approx 2 \cdot \Delta a/a +  \Delta c/c$) to get an approximate value of $\Delta S$, and then use this value and our experimental data to infer the uniaxial pressure dependences. The results are presented in the Table \ref{Tab1}. The estimates for the changes of entropy through the first order phase transitions, evaluated as outlined above, are $\Delta S_{cT} \approx 4$ J/mol K, and $\Delta S_{AFM/ORTHO} \approx 0.3$ J/mol K. This significantly reduced ($\sim 3\%$ of $R\ln2$ per Fe) value of the magnetic entropy  at $T_{AFT/ORTHO}$ (consistent with a very small features in heat capacity observed for these transitions in cobalt-doped BaFe$_2$As$_2$ \cite{nin08b}) indicates reduced magnetic moments on Fe and a dominance of an  itinerant character of the magnetism in these materials.

\begin{table}[ht]
\caption{Relative change of lattice parameters and uniaxial pressure derivatives} 
\centering 
\begin{tabular}{ c |  c |  c |  c | c |  c |  c |  c } 
\hline
sample & transition & $\Delta a/a$ & $\Delta c/c$ & $\Delta V/V$ & $d T_{crit}^{exp}/d P$,  & $d T_{crit}/d p_a$,  &  $d T_{crit}/d p_c$,  \\ 
 &  &  &  &  & K/kbar &   K/kbar &    K/kbar \\ 
\hline\hline
$x = 0$,  &  cT   & $0.75 \cdot 10^{-2}$ & $- 2.8 \cdot 10^{-2}$ &$- 1.3 \cdot 10^{-2}$ &  $\approx 17$ [\onlinecite{tor08a,yuw09a}] &  $\approx - 10$ &  $\approx 37$\\  
as grown  &   &                     &		     &                      &                           &                      & \\  
\hline
$x = 0.027$,  & AFM/ORTHO   &   $- 0.4 \cdot 10^{-2}$ & $1.4 \cdot 10^{-2}$  & $0.6 \cdot 10^{-2}$ &  $\approx -110$ [\onlinecite{gat12a}]  &  $\approx75$ &  $\approx - 260 $ \\               
400$^\circ$C/7 days   &    &                     &		     &                      &                           &                      &                      \\                             
\hline
\end{tabular}
\label{Tab1}
\end{table}

The sign convention for the $\Delta L_i/L_i$ in the table is chosen as $\Delta L_i = L_{i, low} - L_{i, high}$, where low and high are low and high temperature sides of the transition respectively. Even though the inferred values of the uniaxial pressure derivatives might have a sizable, $\sim 20\%$ error bars, these do not obscure the main trends.
\\

The results inferred from thermal expansion measurements for superconducting, antiferromagnetic and structural, and collapsed tetragonal phase transitions for Ca(Fe$_{1-x}$Co$_{x}$)$_2$As$_2$ with different thermal treatment share the same trend. Not only, as  reported in the literature, \cite{gat12a} are the hydrostatic pressure derivative of each of these transitions is very large, approaching  record values for inorganic compounds, but the uniaxial pressure derivatives are extremely high as well, with opposite signs to pressure applied in-plane and along the $c$ - axis. For all three transitions the $c$-axis pressure derivative is dominant, exceeding in its absolute value the in-plane derivative by factor of 3 - 3.5. and exceeding the hydrostatic pressure derivative by factor of 2 - 3. This observation calls for detailed study of elastic properties of these materials, their crystal structure and electronic properties. It should be noted, though, that experimentally, this observation points to the need for extreme attention to details (use of grease, glue, etc.) in sample mounting even for ambient pressure measurements, since even small strains associated with differential thermal contraction on cooling can be significant. \cite{bud13a}

\begin{acknowledgments}

We are indebted to George M. Schmiedeshoff, as well as Wilbur and Olivia Porci, for their help in establishing the dilatometry technique in the Ames Laboratory/Iowa State University Novel Materials and Ground States Group and for much propitious advice.  Work at the Ames Laboratory was supported by the US Department of Energy, Basic Energy Sciences, Division of Materials Sciences and Engineering under Contract No. DE-AC02-07CH11358. S.L.B. acknowledges partial support from the State of Iowa through the Iowa State University.

\end{acknowledgments}

\clearpage

\begin{figure}
\begin{center}
\includegraphics[angle=0,width=100mm]{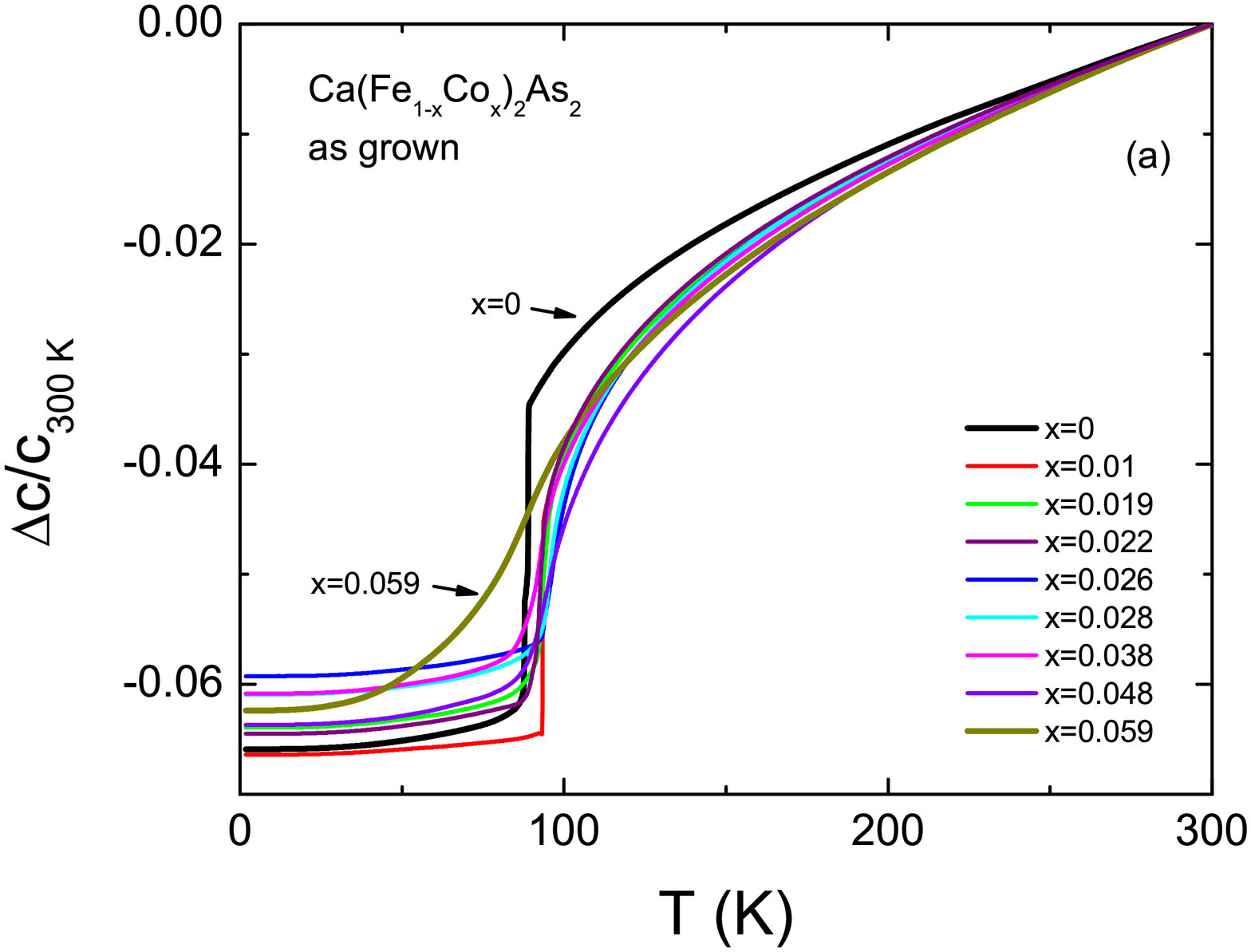}
\includegraphics[angle=0,width=100mm]{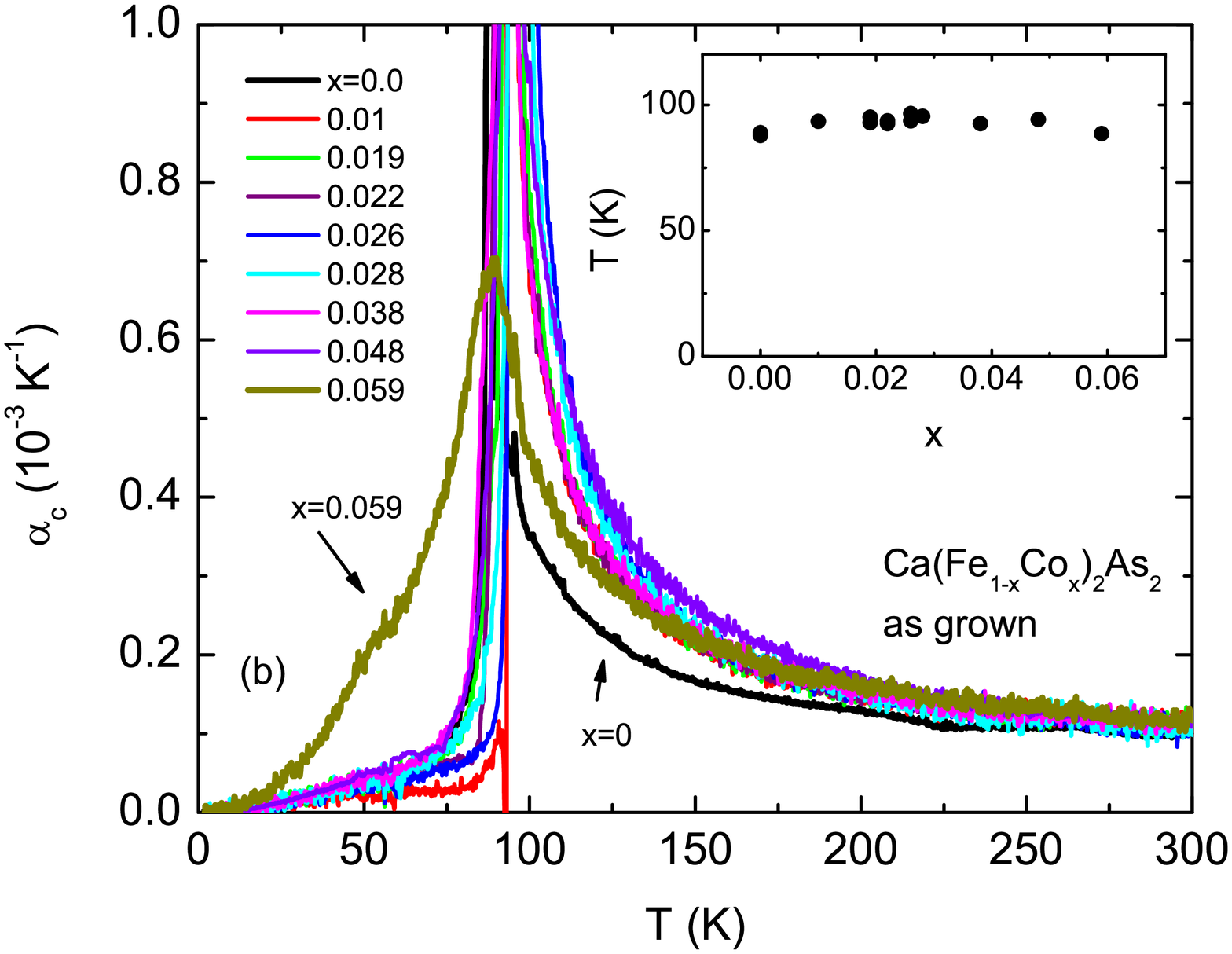}
\end{center}
\caption{(Color online) (a) The temperature-dependent change in $c$-axis dimension normalized with respect to the 300 K value ($\Delta c/c_{300~\textrm{K}} = [c(T) - c(300$ K$)]/c(300$ K$)$) for as-grown Ca(Fe$_{1-x}$Co$_x$)$_2$As$_2$ ($x \leq 0.059$). (b) Temperature-dependent $c$-axis thermal expansion coefficients, $\alpha_c$ for the same samples. Inset: temperature of the cT transition for samples with different cobalt concentrations, as determined by position of the peaks in $\alpha_c$.} \label{F1}
\end{figure}

\clearpage

\clearpage

\clearpage

\clearpage

\clearpage

\clearpage

\clearpage

\clearpage

\begin{thebibliography}{99}

\bibitem{ish09a} Kenji Ishida, Yusuke Nakai, and Hideo Hosono, J. Phys. Soc. Jpn. {\bf 78}, 062001 (2009).

\bibitem{joh10a} D. C. Johnston,  Adv. Phys. {\bf 59}, 803 (2010).

\bibitem{ste11a} G. R. Stewart, Rev. Mod. Phys., {\bf 83}, 1589 (2011).

\bibitem{joh11a} Dirk Johrendt, J. Mater. Chem., {\bf 21}, 13726 (2011).

\bibitem{can10a} Paul C. Canfield and Sergey L. Bud'ko, Annu. Rev. Condens. Matter Phys. {\bf 1}, 27 (2010).

\bibitem{nin11a} N. Ni and Sergey L. Bud'ko, MRS Bull. {\bf 36}, 620 (2011).

\bibitem{man10a} D. Mandrus, A. S. Sefat, M. A. McGuire, B. C. Sales, Chem. Mater. {\bf 22}, 715  (2010).

\bibitem{nin08a} N. Ni, S. Nandi, A. Kreyssig, A. I. Goldman, E. D. Mun, S. L. Bud'ko, and P. C. Canfield, Phys. Rev. B {\bf 78}, 014523 (2008).

\bibitem{can09a} P. C. Canfield, S. L. Bud'ko, N. Ni, A. Kreyssig, A. I. Goldman, R. J. McQueeney, M. S. Torikachvili, D .N. Argyriou, G. Luke, and W. Yu, Physica C {\bf 469}, 404 (2009).

\bibitem{tor08a} Milton S. Torikachvili, Sergey L. Bud'ko, Ni Ni, and Paul C. Canfield, Phys. Rev. Lett. {\bf 101}, 057006 (2008).

\bibitem{kre08a} A. Kreyssig, M. A. Green, Y. Lee, G. D. Samolyuk, P. Zajdel, J. W. Lynn, S. L. Bud'ko, M. S. Torikachvili, N. Ni, S. Nandi, J. B. Le\~ao, S. J. Poulton, D. N. Argyriou, B. N. Harmon, R. J. McQueeney, P. C. Canfield, and A. I. Goldman, Phys. Rev. B {\bf 78}, 184517 (2008).

\bibitem{gol09a} A. I. Goldman, A. Kreyssig, K. Proke\v{s}, D. K. Pratt, D. N. Argyriou, J. W. Lynn, S. Nandi, S. A. J. Kimber, Y. Chen, Y. B. Lee, G. Samolyuk, J. B. Le\~ao, S. J. Poulton, S. L. Bud'ko, N. Ni, P. C. Canfield, B. N. Harmon, and R. J. McQueeney, Phys. Rev. B {\bf 79}, 024513 (2009).

\bibitem{yuw09a} W. Yu, A. A. Aczel, T. J. Williams, S. L. Bud'ko, N. Ni, P. C. Canfield, and G. M. Luke, Phys. Rev. B {\bf 79}, 020511 (2009).

\bibitem{lee09a} Hanoh Lee, Eunsung Park, Tuson Park, V. A. Sidorov, F. Ronning, E. D. Bauer, and J. D. Thompson, Phys. Rev. B {\bf 80}, 024519 (2009).

\bibitem{tor09a} M. S. Torikachvili, S. L. Bud'ko, N. Ni, P. C. Canfield, and S. T. Hannahs, Phys. Rev. B {\bf 80}, 014521 (2009).

\bibitem{pro10a} K. Proke\v{s}, A. Kreyssig, B. Ouladdiaf, D. K. Pratt, N. Ni, S. L. Bud'ko, P. C. Canfield, R. J. McQueeney, D. N. Argyriou, and A. I. Goldman
Phys. Rev. B {\bf 81}, 180506 (2010).

\bibitem{par08a} Tuson Park, Eunsung Park, Hanoh Lee, T. Klimczuk, E. D. Bauer, F. Ronning and J. D. Thompson, J. Phys.: Cond. Mat. {\bf 20}, 322204 (2008).

\bibitem{ran11a} S. Ran, S. L. Bud'ko, D. K. Pratt, A. Kreyssig, M. G. Kim, M. J. Kramer, D. H. Ryan, W. N. Rowan-Weetaluktuk, Y. Furukawa, B. Roy, A. I. Goldman, and P. C. Canfield, Phys. Rev. B {\bf 83}, 144517 (2011).

\bibitem{ran12a} S. Ran, S. L. Bud'ko, W. E. Straszheim, J. Soh, M. G. Kim, A. Kreyssig, A. I. Goldman, and P. C. Canfield. Phys. Rev. B {\bf 85}, 224528 (2012).

\bibitem{gat12a} E. Gati, S. K\"ohler, D. Guterding, B. Wolf, S. Kn\"oner, S. Ran, S. L. Bud'ko, P. C. Canfield, and M. Lang, Phys. Rev. B {\bf 86}, 220511 (2012).

\bibitem{bud10a} Sergey L. Bud'ko, Ni Ni, and Paul C. Canfield, Philos. Mag. {\bf 90}, 1219 (2010).

\bibitem{reb12a} A. Rebello, J. J. Neumeier, Zhaoshun Gao, Yanpeng Qi, and Yanwei Ma, Phys. Rev. B {\bf 86}, 104303 (2012). 

\bibitem{col09a} E. Colombier, S. L. Bud'ko, N. Ni, and P. C. Canfield, Phys. Rev. B {\bf 79}, 224518 (2009).

\bibitem{sch06a} G. M. Schmiedeshoff, A. W. Lounsbury, D. J. Luna, S. J. Tracy, A. J. Schramm, S. W. Tozer, V. F. Correa, S. T. Hannahs, T. P. Murphy, E. C. Palm, A. H. Lacerda, S. L. Bud'ko, P. C. Canfield, J. L. Smith, J. C. Lashley, and J. C. Cooley, Rev. Sci. Instrum. {\bf 77}, 123907 (2006).

\bibitem{bud09a} S. L. Bud'ko, N. Ni, S. Nandi, G. M. Schmiedeshoff, and P. C. Canﬁeld, Phys. Rev. B {\bf 79}, 054525 (2009).

\bibitem{bar99a} T. H. K. Barron and G. K. White, {\it Heat Capacity and Thermal Expansion at Low Temperatures} (Kluwer Academic/Plenum,
New York, 1999).

\bibitem{nin08b} N. Ni, M. E. Tillman, J.-Q. Yan, A. Kracher, S. T. Hannahs, S. L. Bud'ko, and P. C. Canfield, Phys. Rev. B {\bf 78}, 214515 (2008).

\bibitem{bud13a} S. L. Bud'ko, S. Ran, and P. C. Canfield, unpublished.



\end{thebibliography}
\end{document}